\documentclass[11pt,prd,showpacs,showkeys]{revtex4}
\usepackage{epsfig}
\usepackage{amsmath}
\usepackage{epsfig}
\usepackage{graphicx}
\begin{document}

\date{May 9nd, 2007}
\title{\begin{flushright}{}\end{flushright}
Quark masses and mixings in $E_6\times S_3$ : a montecarlo approach}

\author{Stefano Morisi}
\email{stefano.morisi@mi.infn.it}
\affiliation{Dipartimento di Fisica - Universit\'a degli Studi di Milano - Italy\\ 
and\\
Instituto de Fisica Corpuscular (IFIC) - Centro Mixto CSIC-UVEG,
Valencia, Spain
}

\begin{abstract}

Recently in Ref.~\cite{Caravaglios:2005gf} has been proposed a 
GUT model for fermion masses and mixings with spontaneously broken S$_3$ 
discrete flavor symmetry, where S$_3$ is the permutation group of three 
objects. The S$_3$ breaking pattern in the 
quark sector is not studied and need further investigation.
Since in such a model the number of free parameters is greater 
than the number of experimental observables,
an analytical fit of all the parameters is impossible. 
To go forward with the model building and 
to deal with this problem we have used a statistical analysis.
We have found that S$_3$ is totally broken and the up-type quarks matrix 
is approximatively diagonal while down-type quarks matrix
is not symmetric and it is parametrized by three couplings, $g^d,~g_L^d$ and
$g_3^d$. It has been found that $g_L^d$ is slightly smaller than $g^d=1$
and it is of order one, while $g_3^d\sim \lambda^3$ where $\lambda$ is the 
Cabibbo angle. An analytical study of the dependence of $V_{cb}$ and 
$V_{ub}$ from the couplings $g_L^d$ and $g_3^d$ is also presented.

\end{abstract}
\maketitle

\section{Introduction}
In third quantization  \cite{Caravaglios:2002ws} we can explain the 
origin of the fermion families and the origin of the permutation 
symmetry between them. Third quantization is an extension of the quantum field
theory where the fields are treated as well as particles in second 
quantization.
Discrete flavor symmetries, like the group of permutation of three objects 
$S_3$, are very interesting since they 
can naturally explain neutrino mixing angles and mass hierarchies, for instance see  
Ref.~\cite{Harrison:2002er}. Some recent reviews in the study of hierarchical structure of fermion
masses in the Standard Model and its extensions are reported in Ref\,\cite{Fritzsch:1999ee}.
Indications toward grand unification gauge theory (GUT), 
is the phenomenological tendency
to unify of the gauge couplings, and the theoretical implicit possibility to explain
charge quantization and anomaly cancellation.
In Ref.~\cite{Caravaglios:2005gf} 
it has been proposed a model based on  $E_6\times S_3$ 
in order to arrange quark, 
lepton and neutrino masses and mixings.
Minimal GUT models give strong relations between quarks
and leptons Yukawa matrices. For example in minimal SO(10) models up quarks Yukawa
matrix is equal to the Dirac neutrino Yukawa matrix.
However quarks and neutrinos mass and mixing hierarchies are very different.
In fact, neutrino oscillation data show a maximal atmospheric angle, 
large solar angle and $\theta_{13}$ close to zero, see Ref.~\cite{Fogli:2005cq}. 
In contrast
the greatest angle in the quark sector, namely the Cabibbo one,
is smaller
than the corresponding leptonic mixing angle, see 
Ref.~\cite{Charles:2004jd}.   
Phenomenological quark/lepton difference seems in contrast with GUT features.
Even though in $E_6$ models it is possible to 
make a distinction between charged fermions and neutrinos as  showed in Ref.~\cite{Caravaglios:2005gf}.
Authors evidenced this possibility by means of the two standard model singlets contained in $E_6$.
%
%
In such a model 
quarks Yukawas only emerge after
$S_3$  symmetry is broken by not renormalizable dimension five operators, 
while renormalizable operators only contribute to Dirac neutrino Yukawas.
However the study of $S_3$ breaking pattern  in the quark sector has not been accomplished in 
Ref.~\cite{Caravaglios:2005gf} and it will be performed in this paper.  

We will show that in our model  up and down-type quark mass
matrices are parametrized both by five complex couplings and three
real parameters. Since the number of free parameters 
in the quark sector of the model is larger than the number of experimental data,
a direct fit is impossible.
One possibility to deal with this problem, 
is comparing the theoretical expectations with the experimental data by means
of a montecarlo method (see Ref.~\cite{Caravaglios:2002br}) that assigns a probability weight to 
points in the parameter space
of the model according to a goodness of fit criterion. In this way we find the statistically
preferred textures of the mass matrices for up and down-type quarks.
In particular we select a large number of points in the parameters space  
initially uniformed distributed. 
After applying the experimental constrains, not all the selected points agree 
with data. To apply the experimental constrains it is been used  the 
$\chi^2$ function defined as below 
\begin{equation}\label{chi20} 
\chi^2(O^{{\rm t}h})=\sum_{i}\left( \frac{O^{{\rm t}h}_i- O^{{\rm e}xp}_i}{%
\sigma^{{{\rm e}xp}}_i} \right)^2
\end{equation}
where $O^{{\rm e}xp}_i$ and $ \sigma^{{\rm e}xp}_i$ are respectively
the experimental data and their standard deviations while $O^{{\rm t}h}_i$
are the corresponding values predicted by the model.
The points in the configuration space are statistically taken
with a probability proportional to exp$(-\chi^2/2)$. 
After experimental constrains are applied we get a not uniform distribution in the parameters space.
%
Regions very high populated
will be interpreted statistically  more probable than regions with low density population.
This method selects models statistically
preferred by data giving the magnitude of the fitting parameters.\\

The paper is organized as follows: in section~\ref{sec4} we give
the up and down quark mass matrices textures derived by 
$E_6\times S_3$ model, in 
section~\ref{sec5} we introduce the numerical method used and we report
the results, in section~\ref{sec6} we give the conclusions.

\section{S$_3$ symmetry and the quark mass matrices}\label{sec4}
The discrete group of permutation of three objects, called $S_3$, contains six elements and
it has four irreducible representations, namely a doublet {\bf 2},
one symmetric singlet {\bf 1} and one antisymmetric singlet ${\bf 1'}$, see for instance Ref.~\cite{s3e}.
The product of two doublets is 
\begin{equation}\label{eq1}
{\bf 2 \times 2 = 2+1+1'}.
\end{equation}
In particular, if $(\chi^a_1,\chi^a_2)$ and $(\chi^b_1,\chi^b_2)$ are two $S_3$ doublets,
their product is
\begin{eqnarray*}
\chi_D&=&\frac{1}{\sqrt{2}}(\chi^a_2~\chi^b_1+\chi^a_1~\chi^b_2~,~\chi^a_1~\chi^b_1-\chi^a_2~\chi^b_2)\\
\chi_S&=&\frac{1}{\sqrt{2}}(\chi^a_1~\chi^b_1+\chi^a_2~\chi^b_2)\\
\chi_A&=&\frac{1}{\sqrt{2}}(\chi^a_2~\chi^b_1-\chi^a_1~\chi^b_2)
\end{eqnarray*}
where $\chi_D$ is the doublet of eq.(\ref{eq1}) while $\chi_{S}$ and $\chi_{A}$ are respectively the 
symmetric and the antisymmetric singlets.

Let be $\psi_1,~\psi_2$ and $\psi_3$ three fields belonging to a triplet reducible representation {\bf 3} of $S_3$.
Then the triplet reducible representation decompisition is {\bf 3}={\bf 2}+{\bf 1}:
\begin{eqnarray}
\psi_S&=&\frac{1}{\sqrt{3}}(\psi_1+\psi_2+\psi_3)\nonumber\\
\psi_D&=&\left(\frac{1}{\sqrt{2}}(-\psi_2+\psi_3)~,~\frac{1}{\sqrt{6}}(-2~\psi_1+\psi_2+\psi_3)\right)\label{eq1b}
\end{eqnarray}
where $\psi_S$ is a symmetric singlet of $S_3$ and $\psi_D$ is a doublet of $S_3$.

In $E_6\times S_3$ model of Ref.~\cite{Caravaglios:2005gf} 
the quarks Yukawa matrices come from dimension five operators below 
\begin{equation}
L_{Yuk}=\sum_{i\ne j\ne k}\left(g_{ijk}^u~H_{u}\bar{Q}_{Li}u_{Rj}\phi_k^u+g_{ijk}^d~H_{d}\bar{Q}_{Li}d_{Rj}\phi_k^d\right)+{\rm %
h.c.}  \label{eq:1}
\end{equation}
where $i,j,k=1,2,3$, the scalar fields $\phi^{u,d}_i$ are Standard Model 
singlets, $H_u$ and $H_d$ are electroweak doublets, $Q_{Li}$ are
the left-handed fermion electroweak doublets while $u_{Ri}$ and $d_{Ri}$ are 
right-handed fermion electroweak singlets
and $g_{ijk}^{u,d}$ are $S_3$ tensors. 
We assign
the left-handed fermions, the right-handed fermions and the scalar $\phi^{u,d}_i$
to triplet reducible representations of $S_3$, namely
\begin{eqnarray*}\label{eq2}
&&Q_L=(u_i,d_i)_L~\sim~{\bf 3}~,~~
u_{R~i}~\sim~{\bf 3}~,~~d_{R~i}~\sim~{\bf 3},~\phi_i^{u,d}~\sim~{\bf 3}. 
\end{eqnarray*}
The Higgs electroweak doublets are $S_3$ 
symmetric singlets, $H_{u,d}\sim{\bf 1}$.
With this assignment both the operators in eq.~(\ref{eq:1}) transform with respect to the 
permutation group as follows 
\begin{eqnarray}
&&{\bf 3_a\times 3_b \times 3_c\times 1}={\bf (2_a+1_a)\times(2_b+1_b)\times(2_c+1_c)\times 1}=\label{eq4b}\\
&&={\bf (2_a\times 2_b \times 2_c~+~2_a\times 2_b \times 1_c+2_a\times 1_b \times 2_c}+
+{\bf1_a\times 2_b \times 2_c+1_a\times 1_b \times 1_c) \times 1}.\nonumber
\end{eqnarray}
First four terms of five in eq.(\ref{eq4b}) 
contain the symmetric 
and the antisymmetric singlets, see eq.\,(\ref{eq1}). 
Only symmetric singlets give $S_3$ invariant interactions because we have chosen the Higgs
fields in symmetric singlet representation and 
{\bf 1$\times$1$^\prime =$ 1$^\prime$}. 
Consequently Yukawa interactions in eq.\,(\ref{eq:1}) give only five terms $S_3$ invariant 
\footnote{Namely, assuming that the Higgs field belong to the {\bf 1}
symmetric (with respect the Z$_2$ parity symmetry) representation of S$_3$ is equivalent to study the S$_3\times$Z$_2$ group. } 
both in up and down quark sectors.

The lagrangian is, see also Ref.~\cite{s3c},
\begin{eqnarray}
L_{yuk}&=&(c_1^d~\overline{Q}_D d_D \phi_D+c_2^d~\overline{Q}_D d_S \phi_D+c_3^d~\overline{Q}_S d_D \phi_D+c_4^d~\overline{Q}_D d_D \phi_S+c_5^d~\overline{Q}_S d_S \phi_S)~H^d\nonumber\\
&+&(c_1^u~\overline{Q}_D u_D \phi_D+c_2^u~\overline{Q}_D u_S \phi_D+c_3^u~\overline{Q}_S u_D \phi_D+c_4^u~\overline{Q}_D u_D \phi_S+c_5^u~\overline{Q}_S u_S \phi_S)~H^u\nonumber\\
&+&~\mbox{h.c.}\label{yuk1}
\end{eqnarray}
where $Q_D,~d_D,~u_D,~\phi_D,~Q_S,~d_S,~\phi_S$ and $u_S$ follow from relations in eq.(\ref{eq1b})
and $c_i^{u,d}$ are arbitrary complex constants.
With the following redefinition of the complex constants
\begin{equation}\label{sost1}
\begin{array}{ccc}
c_1&=&\sqrt{\frac{2}{3}}~(g - g_2 -g_L - g_R + g_3)\\
c_2&=&\sqrt{\frac{2}{3}}~(g - g_2 +2~g_L - g_R - g_3)\\
c_3&=&\sqrt{\frac{2}{3}}~(g - g_2 -g_L +2~ g_R - g_3)\\
c_4&=&\sqrt{\frac{2}{3}}~(g +2~ g_2 -g_L - g_R - g_3)\\
c_5&=&\frac{1}{\sqrt{3}}~g +2( g_2 +g_L + g_R + g_3)\\
\end{array}
\end{equation}
and using relations in eq.\,(\ref{eq1b}), from eq.~(\ref{yuk1}) we get the following down quarks 
lagrangian  
\begin{equation}
\sum_{i\ne j\ne k} \left(g \bar d_L^i d_R^i H \phi_i +g_2 \bar d_L^i d_R^i H \phi_j+ g_L \bar d_L^i d_R^j H \phi_i+
g_R \bar d_L^i d_R^j H \phi_j+g_3 \bar d_L^i d_R^j H \phi_k\right)+h.c.
\end{equation}
and the corresponding mass matrix is
\begin{eqnarray}\label{M} 
M^{d}&=&
g^{d}v_H^{d}~\left(
\begin{array}{ccc} 
v_1^d & 0&0 \\
0 & v_2^d&0 \\
0 &0 &v_3^d 
\end{array}
\right)+
g_2^{d}v_H^{d}~\left(
\begin{array}{ccc} 
v_2^d+v_3^d & 0&0 \\
0 & v_1^d+v_3^d&0 \\
0 &0 & v_1^d+v_2^d
\end{array}
\right)^{}+\nonumber\\
&+&g_L^{d}v_H^{d}~\left(
\begin{array}{ccc} 
0 &v_1^d &v_1^d \\
v_2^d &0 &v_2^d \\
v_3^d &v_3^d &0 
\end{array}
\right)^{}+
g_R^{u}v_H^{u}~\left(
\begin{array}{ccc} 
0 & v_2^d&v_3^d \\
v_1^d &0 &v_3^d \\
v_1^d &v_2^d &0 
\end{array}
\right)^{}+
g_3^{d}v_H^{d}~\left(
\begin{array}{ccc} 
0 & v_3^d& v_2^d\\
v_3^d &0 & v_1^d\\
v_2^d & v_1^d & 0 
\end{array}
\right)^{}
\end{eqnarray}
where $v_i^d=\langle \phi_i^d \rangle$ and $v_H^d=\langle H^d \rangle$.
We observe that $M^d$ is parametrized  
by five complex coupling constants and three scalar vevs, namely 
13 real parameters.
Similar lagrangian and mass matrix can be obtained for up quarks.

We want to fit the hierarchies of the coupling constants $g_2,~g_L,~g_R$
and $g_3$\footnote{The $|g|$
of the complex constant $g$ can be reabsorbed in $\langle \phi_i \rangle$ 
and its phase is unphysical.}, 
then we parametrize their absolute values as power of the Cabibbo angle as below
%
\begin{equation}\label{ttt}
g_2^{u,d}~\equiv~a_2^{u,d}~\lambda^{t_2^{u,d}}~,~~g_L^{u,d}~\equiv~a_3^{u,d}~\lambda^{t_L^{u,d}}~,~~ g_R^{u,d}~\equiv~a_4^{u,d}~\lambda^{t_R^{u,d}}~,~~g_3^{u,d}~\equiv~a_5^{u,d}~\lambda^{t_3^{u,d}}
\end{equation}
where $\lambda=0.2$, the $t$ exponents are free real parameters and 
$a_2^{u,d}$, $a_3^{u,d}$, $a_4^{u,d}$ and $a_5^{u,d}$ are complex phases.

\section{The numerical fit}\label{sec5}
\subsection{The method}

The goal is to extract the values of $t_{2}^{u,d}$, $t_{R}^{u,d}$, 
$t_{L}^{u,d}$ and $t_{3}^{u,d}$ of eq.~(\ref{ttt}), namely the magnitude order
of the $g$s couplings, and the values 
of the vevs of the
scalars $v_1^{u,d}$, $v_2^{u,d}$ and $v_3^{u,d}$ from the
experimental measurements. A direct fit of the data is not possible, since
the number of free parameters in eq.~(\ref{M}) 
is much larger than the number of observables, six mass
eigenvalues plus four CKM parameters.
The main obstacle comes from the coefficients $a_{i}^{u,d}$, whose phases
are not theoretically known. To cope with them, we will treat
this uncertainty as a theoretical {\it systematic error}. Namely, we have
assigned a flat probability to all the coefficients $a_{i}^{u,d}$ with 
\begin{eqnarray}
&&|a_i^{u,d}|=1~,~~0<{\mathrm{arg}}(a_i^{u,d})<2 \pi.
\end{eqnarray}
The exponents and the vevs are chosen in the ranges
\begin{equation}\label{cutoff}
\begin{array}{c}
0< t_2^{u,d},~t_R^{u,d},~t_L^{u,d},~t_3^{u,d}<8,\\
0<v_1^{u,d},~v_2^{u,d}<v_3^{u,d}=1,
\end{array}
\end{equation}
and we randomly take them with a flat
distribution in logarithmic scale. $v_1^{u,d},~v_2^{u,d}$ and $v_3^{u,d}$ 
must satisfy the
above constraints since (by definition) we choose the entry (3,3) of the
matrices in (\ref{M}) to be the largest one. 
For any random choice of the coefficients $a_{i}^{u,d}$, of the
exponents $t$ and of the vev's $v_i^{u,d}$ we respectively get two
numerical matrices for the up and the down sectors. The
diagonalization of these two matrices, gives  six eigenvalues, corresponding to
the physical up and down quark masses to be compared with the 
experimental values of the masses runned at the unification scale 
$2\cdot 10^{16}Gev$ (see Ref.~\cite{Das:2000uk} ) and reported in Table 
\ref{tab:mass}. 
\begin{table}[h]
{\begin{tabular}{clrc} 
\hline
&Mass& Reference value&\\
\hline
\hspace{4.cm}&$m_u$(MeV)& 0.8351$^{+0.1636}_{-0.1700}$&\hspace{4.cm}\\
&$m_c$(MeV)& 242.6476$^{+23.5536}_{-24.7026}$&\\
&$m_t$(GeV)& 75.4348$^{+9.9647}_{-8.5401}$&\\
&$m_d$(MeV)& 1.7372$^{+0.4846}_{-0.2636}$&\\
&$m_s$(MeV)& 34.5971$^{+4.8857}_{-5.1971}$&\\
&$m_b$(GeV)& 0.9574$^{+0.0037}_{-0.0169}$&\\
\hline
\end{tabular}
\caption{Quark masses runned at $2\cdot 10^{16}Gev$ scale in nonSUSY standard model.}\label{tab:mass}} 
\end{table}
The multiplication of the two 
unitary matrices that diagonalize on the left the up and down
mass matrices respectively, yields the quark mixing CKM matrix. The CKM is parametrized by 
the four Wolfenstein parameters $\lambda$, $A$, $\rho$ and $\eta$ and 
their experimental values (see Ref.~\cite{Charles:2004jd}) are reported in 
Table~\ref{tab:ckm}.
\begin{table}[h]
{\begin{tabular}{cc}
\hline
CKM parameter & Reference value \\
\hline
$\lambda$ & $0.2237\pm 0.0033$ \\ 
$\left | {\rm V}_{cb} \right |$ & $(41.0 \pm 1.6)\times 10^{-3}$ \\ 
$\rho$ & $0.225 \pm 0.038$ \\ 
$\eta $ & $0.317\pm 0.041 $ \\
\hline
\end{tabular} 
\caption{CKM parameters. }\label{tab:ckm}}
\end{table}
We have collected a large statistical
sample of events. Each one event can be compared with the
experimental data through the $\chi^{2}$, namely the event is
accepted with probability 
\begin{equation}
P(O^{{\rm t}h}_i)={e^{-\frac{1}{2}\chi ^{2}(O^{{\rm t}h}_i)}}  \label{chi2}
\end{equation}
where the $\chi^2$ is defined in eq.~(\ref{chi20}).
Before applying the
experimental constraints, events are homogeneously distributed in the
variables $t_i^{u,d},~v_i^{u,d}$, and probability distributions are flat, but
after applying the weight corresponding to eq. (\ref{chi2}), only points
lying in well defined regions of the space $t_i^{u,d}$ and $v_i^{u,d}$  
have a good chance to survive and the events are not uniformally
distributed in the configuaration space.
%
%
The density is related to the probability that an event
predicts masses and mixings in a range compatible with the
experiments. 
%
%
Even if our montecarlo approach gives more predictive and accurate
models, we also emphasize that one should not think these results as
true experimental measurements. They only give ``natural'' range of
values for the exponents $t^{u,d}_i$ and $v^{u,d}_i$.

\subsection{The result}
The statistical analysis gives different regions of points
in the parameters space accordingly with the distribution exp$(-\chi^2/2)$.
Due to a numerical instability  
we are unable to compare different regions of solutions. 
Whatever isolated points in the configuration space are 
statistically 
disfavored since unphysical phases in eq.~(\ref{ttt}) should be tuned in 
order to fit the experimental data. 
While points in regions with high density population are statistically 
preferred by data as explained in the previous section and in 
Ref.~\cite{Caravaglios:2002br}.

\begin{figure}[h]
\begin{center}
\includegraphics*[angle=-90,width=80mm]{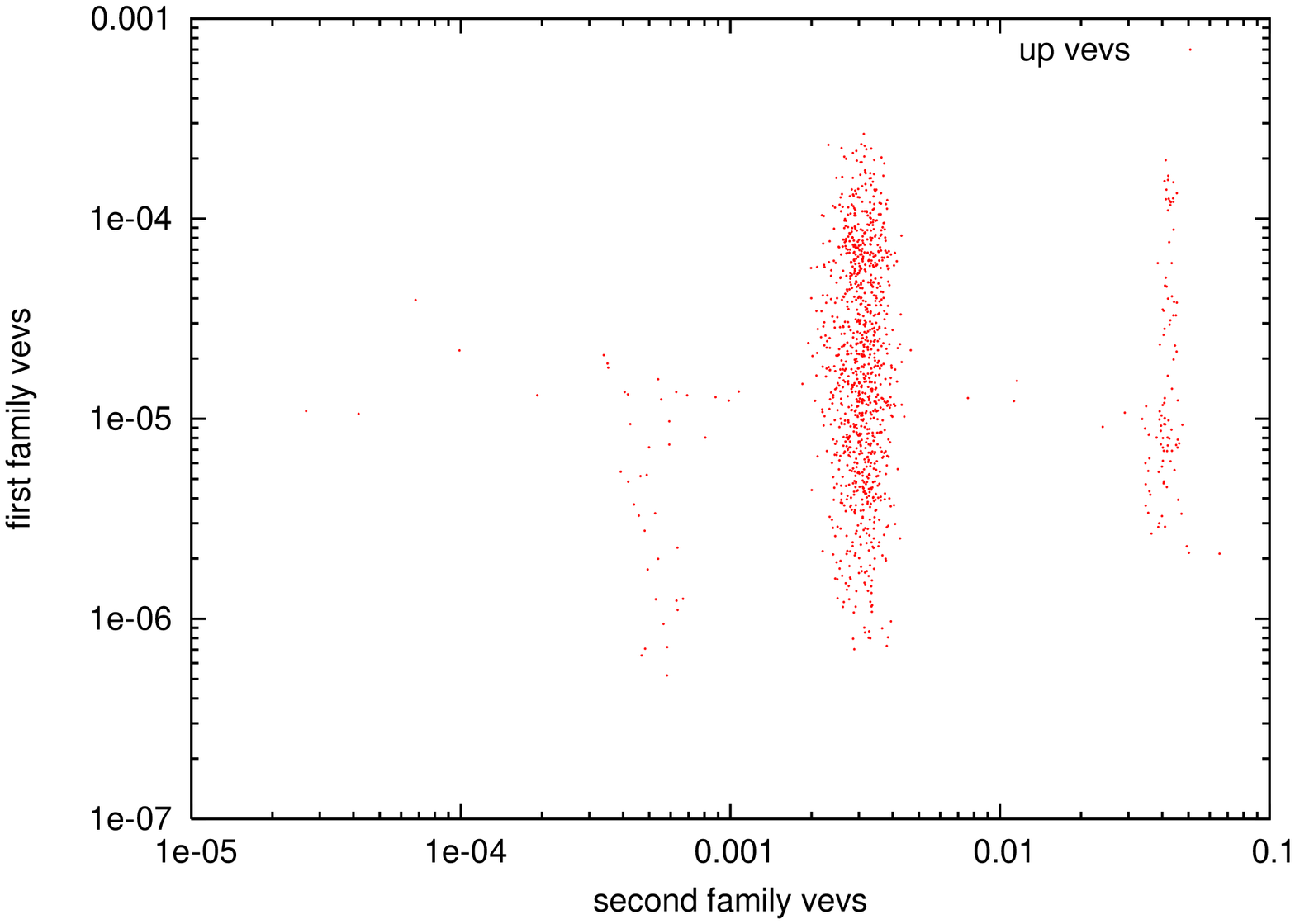}
\includegraphics*[angle=-90,width=80mm]{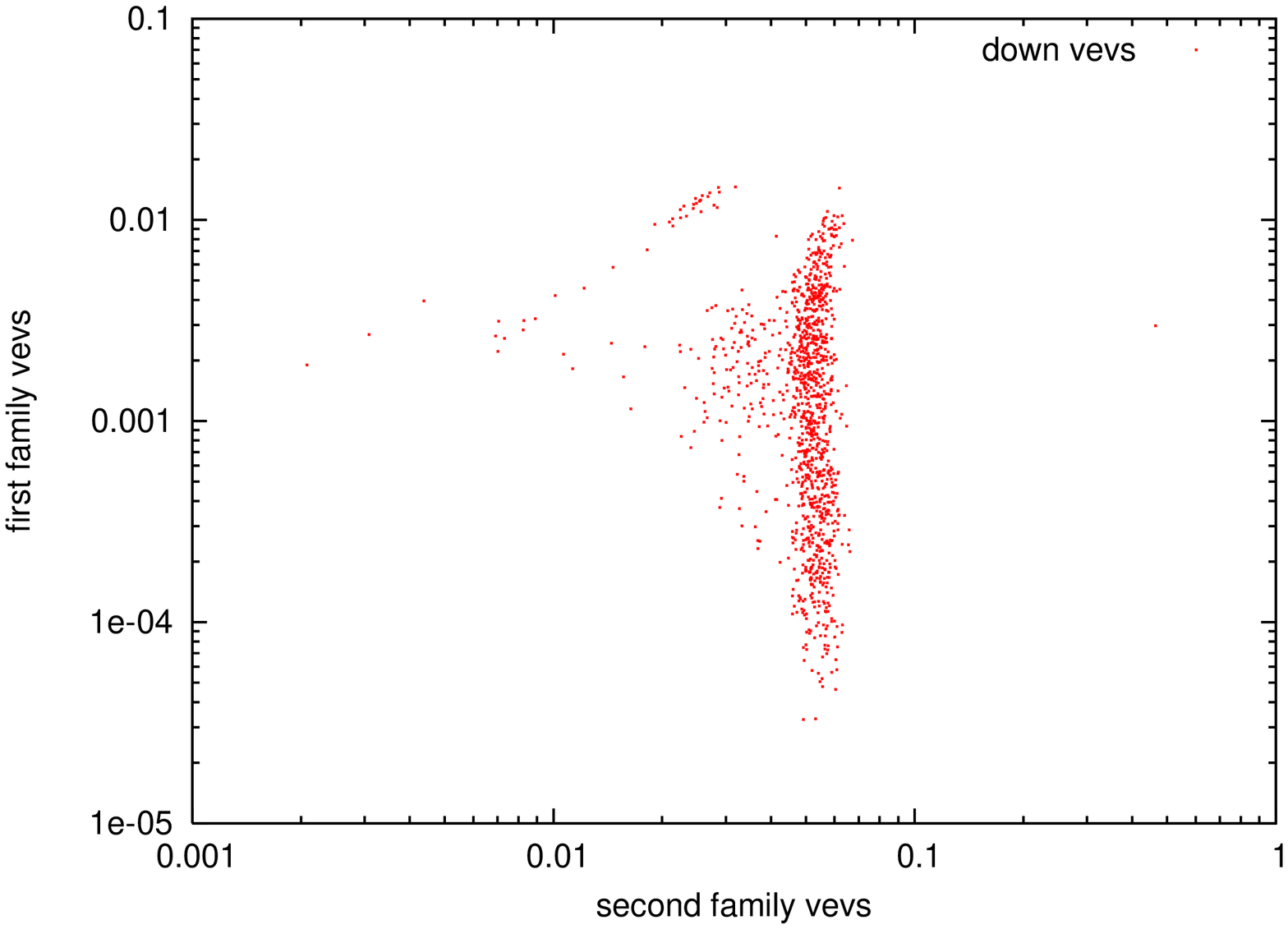}
\end{center}
\caption{
Values of the VEV of the scalar $SU(2)$ 
singlet fields in the up on the {\it left} and down sectors on the {\it right}.}
\label{fig1}
\end{figure}

Let us now to consider the scalar vevs obtained with the montecarlo 
analysis. 
In Fig.~\ref{fig1} are plotted on the left up-type quark vevs
and on the right the down-type quark vevs, in particular it is 
plotted first generation in the $y$ axis {\it vs} second 
generation in the $x$ axis.
The third generation of vevs are fixed to be $v_3^u=1$ and $v_3^d=1$.
From Fig.~\ref{fig1} we have the bounds below
\begin{eqnarray*}
10^{-6}~<~&v_1^u&~<~10^{-4}\\
2\cdot 10^{-3}~<~&v_2^u&~<~3\cdot 10^{-3}\\
10^{-4}~<~&v_1^d&~<~10^{-2}\\
2\cdot 10^{-2}~<~&v_2^d&~<~6\cdot 10^{-2}
\end{eqnarray*}
in agreement with the observed mass hierarchies at the unification scale (see Table~\ref{tab:mass})
\begin{eqnarray}
&&v_1^u:v_2^u:v_3^u=m_u:m_c:m_t=\lambda^7:\lambda^4:1,\label{vu}\\
&&v_1^d:v_2^d:v_3^d=m_d:m_s:m_b=\lambda^4:\lambda^2:1.\label{vd}
\end{eqnarray}
Relations (\ref{vu}) and (\ref{vd}) show that in quark sector the S$_3$ symmetry is totally broken.\\

\begin{figure}[h]
\begin{center}
\includegraphics*[angle=0,width=70mm]{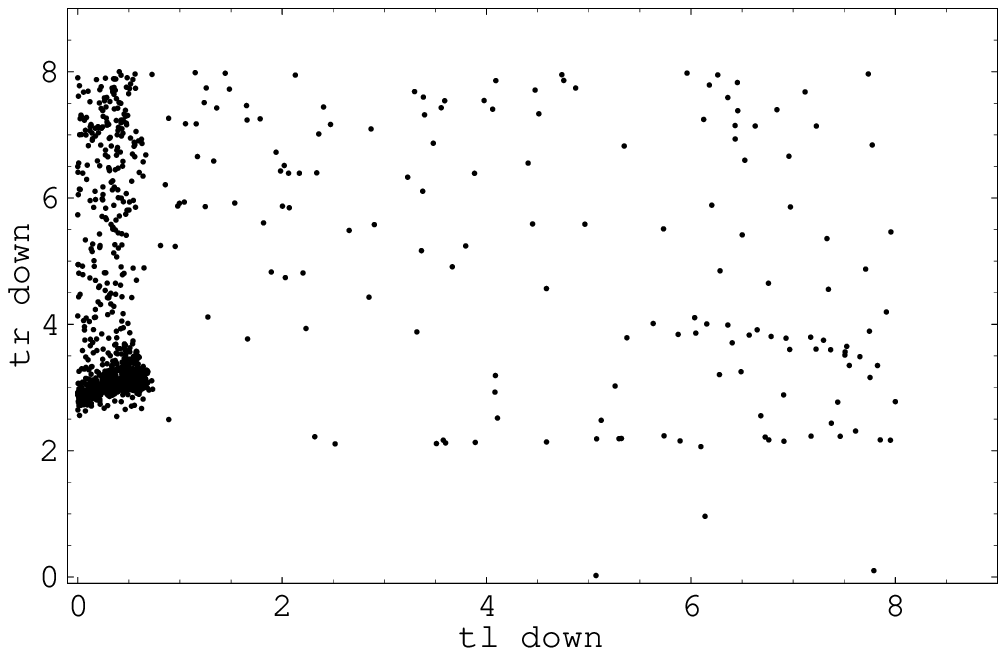}
\includegraphics*[angle=0,width=70mm]{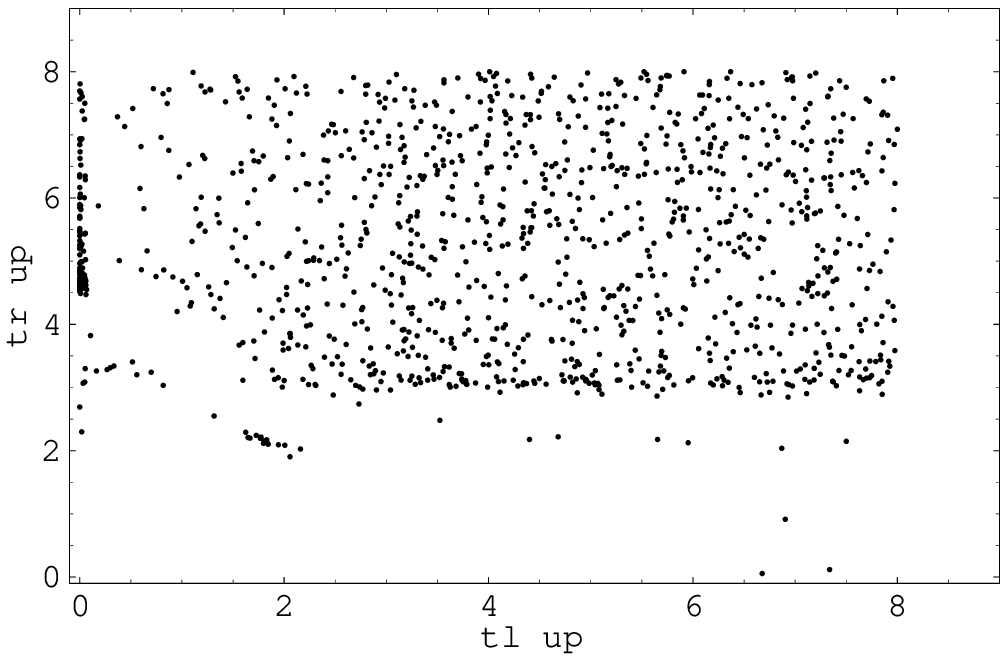}
\includegraphics*[angle=0,width=70mm]{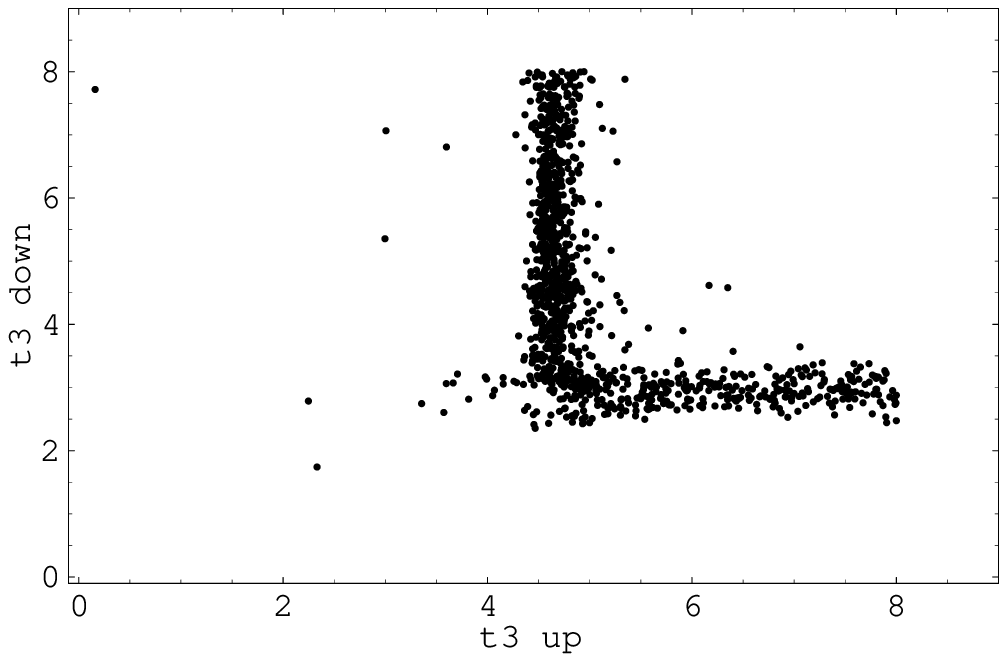}
\includegraphics*[angle=0,width=70mm]{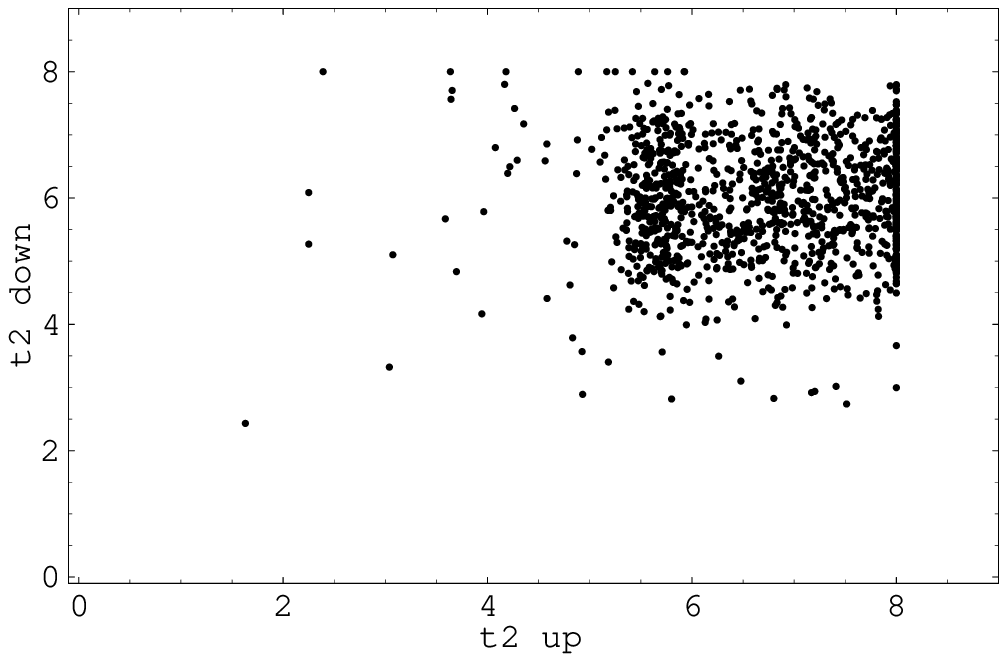}
\end{center}
\caption{$t$ exponents that parametrize the magnitude of the $g$ couplings 
in the general case. We have plotted the plane $(t_L^d,t_R^d)$, 
$(t_L^u,t_R^u)$, $(t_3^u,t_3^d)$ and $(t_2^u,t_2^d)$.}
\label{fig2}
\end{figure}

The magnitude of $g_2^{u,d},~g_L^{u,d},~g_R^{u,d}$ and 
$g_3^{u,d}$ are related to the $t$s exponents like in eq.~(\ref{ttt}).
In Fig.\,(\ref{fig2}) we report the montecarlo results for the
$t$s exponents given by the statistical analysis
allowed after the application of the experimental constrains eq.~(\ref{chi2}). 

From these figures it is not difficult to obtain 
the following bounds for the $t$ exponents 
\begin{equation}\label{texp1}
\begin{array}{ccccccc}
0< &t_L^u& <8  &;&   3< &t_R^u&<8, \\
4.5< &t_3^u& <8  &;&   5<& t_2^u&<8.  \\
\end{array}
\end{equation}
\begin{equation}\label{texp2}
\begin{array}{ccccccc}
0< &t_L^d &<1  &;& 2.5< &t_R^d& < 8,  \\
2.5< &t_3^d& <3 &;& 5< &t_2^d& <8. 
\end{array}
\end{equation}
The upper bounds $t=$8 in eq.~(\ref{texp1}) and eq.~(\ref{texp2})
is the cutoff used in the montecarlo method, see eq.~(\ref{cutoff}).
An upper bound $t<\alpha <8$ correspond  to a $g$ coupling lower
bound, namely $|g|>\lambda^\alpha$.
In the up quark sector, $g$s couplings 
are not bounded from below since the 
corresponding $t$s exponents could be large, namely in  eq.~(\ref{texp1}) 
all the $t$s exponents can be taken equal to
the cutoff used in the statistical 
analysis. This means that all the $g$s couplings in the up quark sector 
can be very small of order $\lambda^8$. 
In contrast, in the down sector the couplings $g_L^d$ and $g_3^d$ 
are bounded from  below\footnote{There are two possibilities, namely the 
up quark mass matrix can  be symmetric (first possibility) or 
antisymmetric (second possibility), while the down quark mass 
matrix is always antisymmetric. The first possibility is in 
agreement with SU(5) unification expectations.
In fact it is not difficult to show that SU(5) invariance implicates 
$g_L^u=g_R^u$.}
\begin{equation} \label{g3g5}
|g_L^d|>\lambda,~~|g_3^d|>\lambda^3,
\end{equation}
while the couplings $g_2^d$ and $g_R^d$ in the down sector 
can be very small like in the up quark sector.
Experimental data statistically prefer models where the absolute value of 
the $|g_L^d|$ coupling is large, see eq.~(\ref{g3g5}). In 
particular $|g_L^d|$  can be close to one or to the maximal 
value allowed in the montecarlo analysis, see eq.~(\ref{cutoff})
and it is slightly smaller than $g^d=1$. 
Indeed from relations~(\ref{texp2}) and  (\ref{g3g5}) the $g_3^d$
coupling also has an upper bound, that is $|g_3^d|< \lambda^{2.5}\ll \lambda$.
Therefore from the statistical analysis we get the following hierarchies
between the $g$s couplings
\begin{equation} \label{gh}
1~=~g^u~=~g^d~\simeq~|g_L^d| ~\gg ~|g_3^d|~ \gg ~|g_2^d|
,~|g_R^d|,~|g_2^u|,~|g_L^u|,~|g_R^u|,~|g_3^u|.
\end{equation}
In Table~\ref{tab2} we report a numerical example of values for the $g$s 
coupling and for the vevs that fit good data.

\begin{center}
\begin{table}[h]
\begin{tabular}{ll}
\hline
up&down\\
\hline
&\\
$\phi^u_1/\phi^u_3=m_u/m_t$&$\phi^d_1/\phi^d_3=0.000077$\\
$\phi^u_2/\phi^u_3=m_c/m_t$&$\phi^d_2/\phi^d_3=0.054783$\\
$g=1$ & $g=0.848$ \\
$g_L\simeq 0$ & $g_L=0.305~e^{-i~0.02}$\\
$g_R \simeq 0$ & $g_R=0.002~e^{-i~0.12}$\\
$g_L \simeq 0$ & $g_L~=0.009~e^{i~1.82}$\\
\hline
\end{tabular}
\caption{Example of set of values of $g$ couplings and vevs that fit the data. 
}\label{tab2}
\end{table}
\end{center}

\subsection{The CKM and the couplings $g_L^d$ and $g_3^d$}
The statistical analysis results in previous 
section, show that 
in first approximation the experimental data can be fitted only with
couplings  $g^u,~g^d,~g_L^d$ and $g_3^d$ couplings ($g^{u,d}$ are fixed), besides the vevs of 
the scalar fields. In this section we want to study the relation between
the  $g_L^d$ and $g_3^d$ and the entries of the CKM mixing matrix.
Accordingly we fix the values of the scalar vevs like in eq.~(\ref{vu}) and 
eq.~(\ref{vd}). 
From eq.~(\ref{vu}), eq.~(\ref{vd}) and eq.~(\ref{gh})
we get, up to correction of order $\lambda^8$, the mass matrices below 
\begin{equation}\label{Mup1} 
M^{u}\simeq v_H^u\left(\begin{array}{ccc} 
\lambda^7 &0 & 0 \\
0 &\lambda^4 & 0 \\
0&0 & 1 
\end{array}\right)~,~~~
M^{d}\simeq v_H^d
\left(\begin{array}{ccc} 
 \lambda^4& g_L^d~\lambda^4+g_3^d&g_L~\lambda^4+g_3^d~\lambda^2 \\
g_L^d~\lambda^2+g_3^d & \lambda^2 & g_L^d~\lambda^2+g_3^d~\lambda^4\\
g_L^d+g_3^d~\lambda^2&g_L^d+g_3^d~\lambda^4& 1 
\end{array}\right).
\end{equation}
The up quark mass matrix in eq.~(\ref{Mup1}) is almost diagonal and the 
quark mixing 
matrix $V_{CKM}$ is approximatively given by the unitary matrix $U_d$ that 
diagonalize on the left the down 
quark mass matrix eq.~(\ref{Mup1}), namely $U_d^\dagger~M_d~M_d^\dagger~U_d=D_d^2$.
We observe that in general the $g_L^d$ and $g_3^d$
are arbitrary complex variables. In order to simplify the problem, we assume $g_L^d$ 
and $g_3^d$ reals and the CKM complex phase $\eta$ is not fitted. 

In the standard parametrization
$V_{CKM}=R_{23}~R_{13}~R_{12}$
where $R_{ij}$ are rotations in the $(ij)$ plane of $\theta_{ij}$ angles,
then $|V_{cb}|\approx \sin\theta_{23}$ and $|V_{ub}|\approx \sin\theta_{13}$.
From mass matrices (\ref{Mup1}) we have
\begin{equation}\label{Vcb} 
|V_{cb}|\simeq 
\frac{1}{\sqrt{2}}\sqrt{1+\frac{(1+2 g_L^2)(\lambda^4-1)}{\sqrt{16 g_L^3\lambda^4+(\lambda^4-1)^2+4 g_L^4 (1-\lambda^4+\lambda^8)}}}
\end{equation}
where $g_L=\lambda^{t_L}$, so in first 
approximation $|V_{cb}|$ is only a function of the coupling $g_L^d$.
\begin{figure}[h]
\begin{center}
\includegraphics*[angle=0,width=65mm]{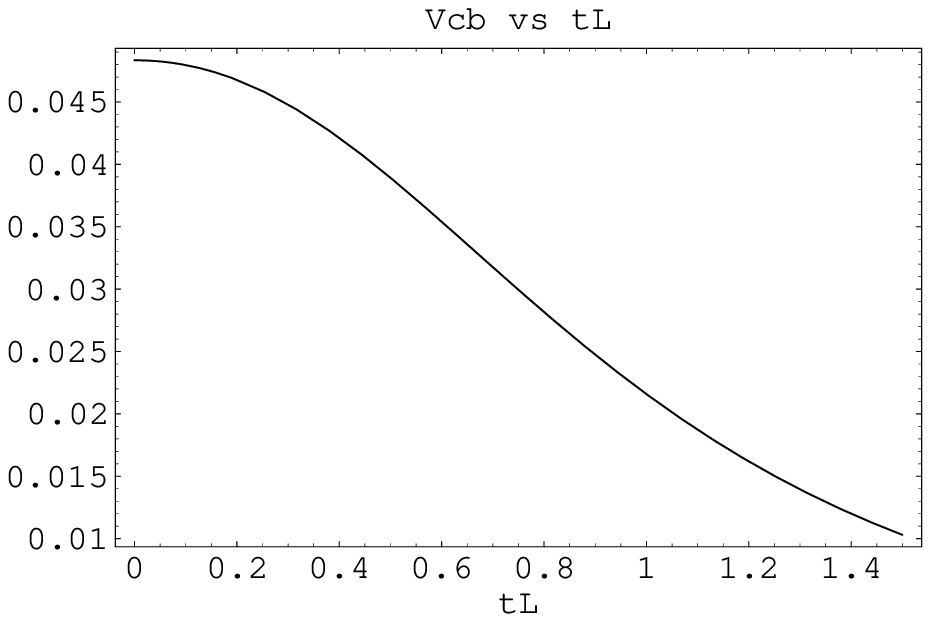}
\includegraphics*[angle=0,width=65mm]{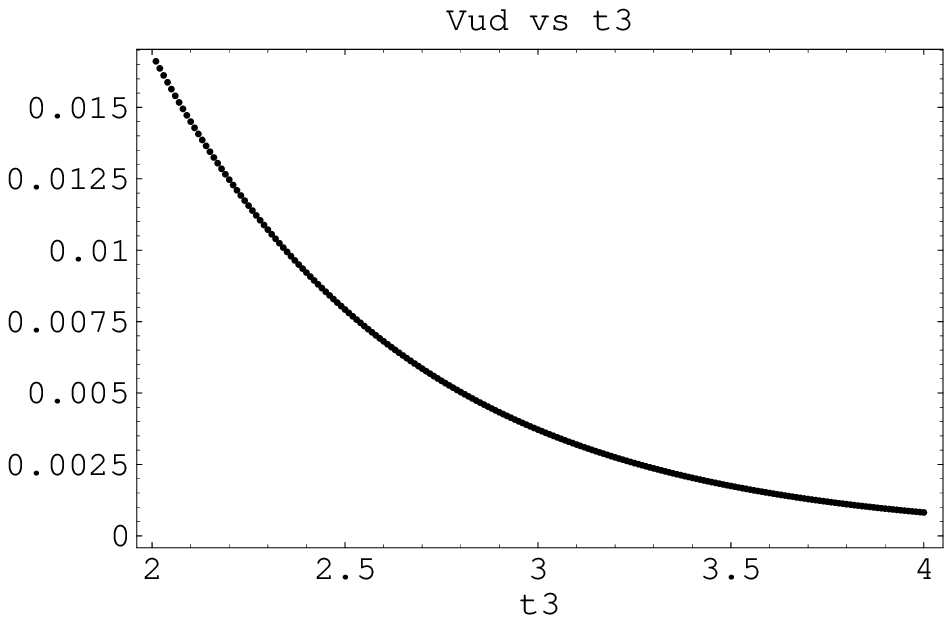}
\end{center}
\caption{Plot of the function $|V_{cb}|=V_{cb}(t_L)$ and $|V_{ub}|=V_{ub}(t_3)$.}
\label{fig8}
\end{figure}
In Fig.~\ref{fig8} we plot the function $|V_{cb}|=V_{cb}(t_L)$. 
We observe that when $t_L\simeq 0.5$ the  corresponding $|V_{cb}|$ 
agree with data and 
$t_L\simeq 0.5$ is close to the value 
from the montecarlo, see in eq.~(\ref{texp2}).

Analogously we can use the remaining free parameter $g_3^d$ 
to fit the $|V_{ub}|$ as below
\begin{equation}\label{Vub} 
|V_{ub}|\simeq
\frac{1}{\sqrt{2}}\sqrt{1+\frac{g_3^2-4 g_3 \lambda^2-3}{\sqrt{
9+g_3^4+24 g_3 \lambda^2-8 g_3^3+2 g_3^2 (10 \lambda^4+4 \lambda^2-1)
}}}.
\end{equation}
In Fig.~\ref{fig8} we have plotted the function $V_{ub}(t_3)$. 
$|V_{ub}|$ agrees with experimental data when $t_3\simeq 3$, 
in agreement with statistical analysis. 
$|V_{ub}|$ is approximatively only a function  of  $g_L^d$ coupling. 
Assuming for instance $t_L=0.55$ and $t_3=3.1$ in eq.~(\ref{Mup1}), we have 
$|V_{us}|\simeq 0.21$, $|V_{ub}|\simeq 0.004$ and $|V_{cb}|\simeq 0.043$ 
that agree with data.

\section{Conclusions}\label{sec6}

Recently it has been proposed in  Ref.~\cite{Caravaglios:2005gf} 
a model based on $E_6\times S_3$ in order to explain fermions mass and mixing hierarchies.
However a full study of $S_3$ breaking pattern was not accomplished.  We carry out 
such a study. 
In the model of  Ref.~\cite{Caravaglios:2005gf} quark mass terms appear as higher order dimension
five operators after $S_3$ is broken. 
We have showed that in such a model quark mass matrices 
are parametrized by five complex couplings 
$g^{u,d},~g_2^{u,d},~g_L^{u,d},~g_R^{u,d},~g_3^{u,d}$
and three real parameters
$v_1^{u,d},~v_2^{u,d},~v_3^{u,d}$
that are the vevs of the scalar fields that break $S_3$. 
Even if we can assume $g^{u}=g^{d}=1$ and $v_1^{u}=v_1^{d}=1$,
the number of free parameters in the model is larger
than the number of experimental data then it is not possible to fit the parameters. 
To go forward we have used a numerical approach 
finding the 
statistically preferred textures of the mass matrices according to a goodness of fit criterion.
We have found that the permutation symmetry $S_3$ is totally broken
and vev hierarchies should be 
$v_1^{u,d}\ll v_2^{u,d}\ll v_3^{u,d}$, and
data statistically prefer solutions with large $|g_L^d|$ coupling, namely
slightly smaller than  $g^{u}=1$, and  $|g_3^d|\approx \lambda^3$ while
the remain couplings in up and down quark textures, 
are negligible and are of order $\lambda^8$. The resulting up quark mass 
matrix is almost diagonal and the quarks mixing matrix is approximatively
given by the down quark mass matrix that is not symmetric. The hierarchies
between the yukawa couplings is 
\begin{equation}\label{conc}
1~=~g^u~=~g^d~\simeq~|g_L^d| ~\gg ~|g_3^d|~ \gg ~|g_2^d|
,~|g_R^d|,~|g_2^u|,~|g_L^u|,~|g_R^u|,~|g_3^u|.
\end{equation}
Ultimately we have studied the relation between the couplings $g_L^d$, $g_3^d$
and the CKM mixing matrix. We have found that 
$V_{cb}$  is approximatively a function of $g_L^d$ coupling and
$V_{ub}$ is approximatively a function of $g_3^d$ coupling.\\


The study of the breaking pattern of the $S_3$ symmetry in the quark 
sector, gives indications to go forward in the model
building. In fact recently in Ref.~\cite{Caravaglios:2006aq} it has been 
studied a model that could explains quark, lepton and neutrino
mixings and masses, making use of the montecarlo
statistical analysis results reported in this paper. 
In particular
in the model of Ref.~\cite{Caravaglios:2006aq} are explained 
the hierarchies in eq.~(\ref{conc}) between the $g$s Yukawa couplings.
In third quantization models it is possible to extend 
the concept of family to the gauge bosons
through the semidirect product of a gauge group $G$, with the 
group of the permutation of $n$ objects, namely 
$G^n>{\hspace{-8pt}}\triangleleft S_n$. In Ref.~\cite{Caravaglios:2006aq}
it has been proposed a $E_6^4 >{\hspace{-8pt}}\triangleleft S_4$ model 
where $E_6$ is the grand unified gauge 
group and $S_4$ is the permutation symmetry of four objects.
It is showed that embedding $S_3\subset S_4$
it is possible to better explain neutrino oscillation, while $E_6^4$
could be useful in order to understand the hierarchies between the $g$s 
couplings reported in eq.~(\ref{conc}).
It is presented a possible breaking pattern
of $E_6^4 >{\hspace{-8pt}}\triangleleft S_4$ in agreement with our montecarlo 
results. In the following we report such a breaking pattern.

Assume that the starting group 
$E_6^4 >{\hspace{-8pt}}\triangleleft S_4$ is broken into 
the group $G_1=(SM\times U(1)_r\times U(1)_t)^4 >{\hspace{-6pt}}\triangleleft S_4$ 
where 
$U(1)_r$ and $U(1)_t$ are defined as
$E_6\supset SO(10) \times U(1)_t 
\supset SU(5) \times U(1) _r\times U(1)_t$ and $SM$ is the Standard Model 
gauge group. 
At some scale $M_1$ the group  $G_1$ breaks into the group 
$G_2=(SM\times U(1)_r\times U(1)_t)^3 >{\hspace{-6pt}}\triangleleft~S_3$.
It is easy to see that  the only operators compatible with $G_2$ are 
$$g ~\bar d_L^i d_R^i H \phi_i.$$
The coupling $g$ will be of order of magnitude $g\simeq M_1/\Lambda$ where 
$\Lambda$ is the characteristic scale of 
$E_6^4 >{\hspace{-8pt}}\triangleleft S_4$.
At another scale $M_2<M_1$ the group $G_2$ breaks into 
$G_3=(SM\times U(1)_{\mathrm{down}})^3 >{\hspace{-6pt}}\triangleleft S_3$
where $U(1)_{\mathrm{down}}$ is defined as the linear combination of $U(1)_r$ 
and $U(1)_t$ so that right-handed down quarks do not carry $U(1)_{\mathrm{down}}$
charge. At this scale also the operators
$$
\sum_{i,j} g_L~ \bar d_L^i d_R^j H^i \phi_i 
$$
are allowed and  $g_L\simeq M_2/\Lambda$. If $M_1\sim M_2$ it is possible
to explain the relation $g\simeq g_L$ in eq.~(\ref{conc}).
At the scale $M_2$ the operators proportional to $g_2^d$, $g_3^d$ and $g_R^d$ 
are not allowed. They only appear after the breaking of 
$U(1)_{\mathrm{down}}$ at some scale $M_3<M_2$. 
If $M_2\gg M_3$ we can explain
the hierarchy $g_L^d\gg g_3^d,~g_R^d,~g_2^d $.

\section*{Acknowledgments}
I would like to thank F. Caravaglios for the helpful discussion and 
G. Altarelli for the useful suggestions.


\begin{thebibliography}{0}    


\bibitem{Caravaglios:2005gf}
  F.~Caravaglios and S.~Morisi,
  arXiv:hep-ph/0510321.
\bibitem{Caravaglios:2002ws}
  F.~Caravaglios,
  arXiv:hep-ph/0211183;
  F.~Caravaglios,
  arXiv:hep-ph/0211129;
  V.~A.~Rubakov,
  Phys.\ Lett.\ B {\bf 214}, 503 (1988);
  M.~McGuigan,
  Phys.\ Rev.\ D {\bf 38}, 3031 (1988);
  S.~B.~Giddings and A.~Strominger,
  Nucl.\ Phys.\ B {\bf 321}, 481 (1989).


\bibitem{Harrison:2002er}
  P.~F.~Harrison, D.~H.~Perkins and W.~G.~Scott,
  Phys.\ Lett.\ B {\bf 530}, 167 (2002)
  [arXiv:hep-ph/0202074];
  K.~S.~Babu, E.~Ma and J.~W.~F.~Valle,
  Phys.\ Lett.\  B {\bf 552}, 207 (2003)
  [arXiv:hep-ph/0206292];
  G.~Altarelli and F.~Feruglio,
  Nucl.\ Phys.\  B {\bf 741}, 215 (2006)
  [arXiv:hep-ph/0512103];
  W.~Grimus and L.~Lavoura,
  JHEP {\bf 0508}, 013 (2005)
  [arXiv:hep-ph/0504153].


\bibitem{Fritzsch:1999ee}
  H.~Fritzsch and Z.~z.~Xing,
  Prog.\ Part.\ Nucl.\ Phys.\  {\bf 45}, 1 (2000)
  [arXiv:hep-ph/9912358];
  G.~Altarelli and F.~Feruglio,
  New J.\ Phys.\  {\bf 6}, 106 (2004)
  [arXiv:hep-ph/0405048];
  C.~H.~Albright and M.~C.~Chen,
  Phys.\ Rev.\  D {\bf 74}, 113006 (2006)
  [arXiv:hep-ph/0608137];
  G.~Altarelli,
{\it In the Proceedings of IPM School and Conference on Lepton and Hadron Physics (IPM-LHP06), Tehran, Iran, 15-20 May 2006, pp 0001}
  [arXiv:hep-ph/0610164].


\bibitem{Fogli:2005cq}
  G.~L.~Fogli, E.~Lisi, A.~Marrone and A.~Palazzo,
  Prog.\ Part.\ Nucl.\ Phys.\  {\bf 57}, 742 (2006)
  [arXiv:hep-ph/0506083];
  M.~Maltoni, T.~Schwetz, M.~A.~Tortola and J.~W.~F.~Valle,
  New J.\ Phys.\  {\bf 6}, 122 (2004)
  [arXiv:hep-ph/0405172];
  J.~N.~Bahcall and C.~Pena-Garay,
  JHEP {\bf 0311}, 004 (2003)
  [arXiv:hep-ph/0305159].


\bibitem{Charles:2004jd}
  J.~Charles {\it et al.}  [CKMfitter Group],
  Eur.\ Phys.\ J.\  C {\bf 41}, 1 (2005)
  [arXiv:hep-ph/0406184];
M.Ciuchini {\it et al.}, JHEP 0107(2001)013.


\bibitem{Caravaglios:2002br}
  F.~Caravaglios, P.~Roudeau and A.~Stocchi,
  Nucl.\ Phys.\ B {\bf 633}, 193 (2002)
  [arXiv:hep-ph/0202055].


\bibitem{s3e}
  E.~Ma,
  New J.\ Phys.\  {\bf 6}, 104 (2004);


\bibitem{s3c}
  J.~Kubo, A.~Mondragon, M.~Mondragon and E.~Rodriguez-Jauregui,
  Prog.\ Theor.\ Phys.\  {\bf 109}, 795 (2003)
  [Erratum-ibid.\  {\bf 114}, 287 (2005)]
  [arXiv:hep-ph/0302196];
  J.~Kubo, A.~Mondragon, M.~Mondragon, E.~Rodriguez-Jauregui, O.~Felix-Beltran and E.~Peinado,
  J.\ Phys.\ Conf.\ Ser.\  {\bf 18} (2005) 380;
  O.~Felix, A.~Mondragon, M.~Mondragon and E.~Peinado,
  arXiv:hep-ph/0610061.

\bibitem{Das:2000uk}
  C.~R.~Das and M.~K.~Parida,
  Eur.\ Phys.\ J.\ C {\bf 20}, 121 (2001)
  [arXiv:hep-ph/0010004].


\bibitem{Caravaglios:2006aq}
  F.~Caravaglios and S.~Morisi,
  arXiv:hep-ph/0611078.




\end{thebibliography}
\end{document}